\documentclass[preprint2]{aastex}
\usepackage{natbib}
\usepackage{epsfig}
\usepackage{amsmath}
\begin{document}

\title{How complex is the obscuration in AGN? New clues from the \textit{Suzaku} monitoring of the X-ray absorbers in NGC~7582}

\author{Stefano Bianchi}
\affil{Dipartimento di Fisica, Universit\`a degli Studi Roma Tre, via della Vasca Navale 84, 00146 Roma, Italy}
\author{Enrico Piconcelli}
\affil{Osservatorio Astronomico di Roma (INAF), Via Frascati 33, I-00040 Monte Porzio Catone, Italy}
\author{Marco Chiaberge}
\affil{Space Telescope Science Institute, 3700 San Martin Drive, Baltimore, MD 21218}
\affil{INAF - IRA, Via P. Gobetti 101, I-40129 Bologna, Italy}
\author{Elena Jimenez Bail\'on}
\affil{Instituto de Astronom\'ia, Universidad Nacional Aut\'onoma de M\'exico, Apartado Postal 70-264, 04510 Mexico DF, Mexico}
\affil{LAEFF Apd. 78 Villanueva de la Ca\~nada - 28691-Madrid, Spain}
\author{Giorgio Matt}
\affil{Dipartimento di Fisica, Universit\`a degli Studi Roma Tre, via della Vasca Navale 84, 00146 Roma, Italy}
\author{Fabrizio Fiore}
\affil{Osservatorio Astronomico di Roma (INAF), Via Frascati 33, I-00040 Monte Porzio Catone, Italy}

\begin{abstract}
We present the results of a \textit{Suzaku} monitoring campaign of the Seyfert 2 galaxy, NGC~7582. The source is characterized by very rapid (on timescales even lower than a day) changes of the column density of an inner absorber, together with the presence of constant components arising as reprocessing from a Compton-thick material. The best fitting scenario implies important modifications to the zeroth order view of Unified Models. While the existence of a pc-scale torus is needed in order to produce a constant Compton reflection component and an iron K$\alpha$ emission line, in this Seyfert 2 galaxy this is not viewed along the line of sight. On the other hand, the absorption of the primary continuum is due to another material, much closer to the BH, roughly at the distance of the BLR, which can produce the observed rapid spectral variability. On top of that, the constant presence of a $10^{22}$ cm$^{-2}$ column density can be ascribed to the presence of a dust lane, extended on a galactic scale, as previously confirmed by \textit{Chandra}. There is now mounting evidence that complexity in the obscuration of AGN may be the rule rather than the exception. We therefore propose to modify the Unification Model, adding to the torus the presence of two further absorbers/emitters. Their combination along the line of sight can reproduce all the observed phenomenology.
\end{abstract}


\section{Introduction}

The presence of absorbing material along the line of sight is generally believed to be the only difference between Type 2 and a Type 1 Active Galactic Nuclei (AGN). This material obscures both the emission lines from the Broad Line Region (BLR) and the X-ray spectrum, being the main ingredient of the so-called Unification Model. It is usually envisaged as a compact `torus', located at a pc scale distance from the nucleus \citep[e.g.][]{antonucci93}. This distance is basically confirmed both by indirect techniques, such as considerations based on photoionization codes \citep[e.g.][]{bmi01,mass06}, and direct `imaging' of the torus itself \citep[e.g.][]{jaffe04}.

However, there is evidence that this simple scenario may not hold for all objects. The co-existence of a Compton-thick torus and a Compton-thin material, extended on a much larger scale, seems to better account for the observed phenomenology \citep[e.g.][]{matt00b}. The latter absorber may be naturally associated to dust-lanes \citep[e.g.][]{mvr98}, or to molecular gas in the galactic disks \citep{lamastra06}. The presence of obscuring matter on large (pc-kpc) scales and detached from the nuclear torus is also supported by \textit{Spitzer} studies of MIR luminous quasars at high z which are very likely hosted by dusty galaxies \citep[e.g.][]{poll08,ms06}

Moreover, \citet{risa02b} showed that a large number of Seyfert 2s presents significant variability of the absorbing column density ($\mathrm{N_H}$) on timescales as low as months, thus suggesting that the absorbing material should be much closer to the nucleus than assumed for the torus, possibly in the BLR itself. This picture seems the only tenable for the objects, which present the most rapid $\mathrm{N_H}$ variations ever observed, within days or even hours: NGC~4388 \citep{elvis04}, NGC~1365 \citep{ris05} and NGC~4151 \citep{puc07}. Is this the end of the torus paradigm? Or is it only an exception on a handful of peculiar objects? 

NGC~7582 (z=0.0053), being included in the \citet{pic82} catalog, has been targeted by most X-ray telescopes: \textit{Einstein} \citep{mp81}, \textit{EXOSAT} \citep{tp89}, \textit{Ginga} \citep{war93}, \textit{ASCA} \citep{schac98,xue98}. The picture that emerged from these studies was that of a flat X-ray spectrum dominated by heavy obscuration. Thanks to the BeppoSAX broad bandpass, \citet{turn00} reported for the first time the detection of a more complex geometry of the absorbing material, likely constituted by two different components, one of which Compton-thick. This scenario was confirmed by a combined imaging analysis performed with \textit{Chandra} and \textit{HST}, which suggested that the Compton-thick torus coexists with a large-scale Compton-thin material associated with the dust lane and circumnuclear gas is photoionized by the AGN along torus-free lines of sight \citep{bianchi07b}.

The most interesting results came from the analysis of the two XMM-\textit{Newton} observations, taken 4 years apart, in 2001 and 2005 \citep{pico07}. Both clearly show a completely different spectral and flux state with respect to the 1998 BeppoSAX observation. The XMM-\textit{Newton} spectrum can be well described by a model consisting of a combination of a heavily absorbed ($\mathrm{N_H}\sim10^{24}$ cm$^{-2}$) power law and a pure reflection component, both obscured by a column density of few $\times10^{22}$ cm$^{-2}$. Notably, \citet{pico07} detect a significant increase by a factor $\sim2$ in the column density of the inner, thicker absorber covering the primary X-ray source, between 2001 and 2005.

In this paper, we present a \textit{Suzaku} monitoring campaign of NGC~7582, which, together with a new XMM-\textit{Newton} observation, confirms the variability of the column density of the inner absorber, but down to timescales smaller than a day.

\section{Observations and data reduction}

\subsection{\textit{Suzaku}}

During the second \textit{Suzaku} Announcement of Opportunity (AO2), we proposed a strategy to observe NGC~7582 at different timescales, from 1 week to about 6 months, allowing us to probe distances as close as the BLR and almost as far as the traditional torus. Moreover, this campaign complemented the scales of the order of years already tested with XMM-\textit{Newton}. Therefore, NGC~7582 was observed four times by \textit{Suzaku} in 2007 (PI: M. Chiaberge): on May 1st and 28th, and November 9th and 16th. X-ray Imaging Spectrometer (XIS) and Hard X-ray Detector (HXD) event files were reprocessed with the latest calibration files available (2008-07-09 release), using \textsc{ftools} 6.5 and Suzaku software Version 9, adopting standard filtering procedures. Source and background spectra for all the three XIS detectors were extracted from circular regions of 2.9 arcmin radius, avoiding the calibration sources. Response matrices and ancillary response files were generated using \textsc{xisrmfgen} and \textsc{xissimarfgen}. We downloaded the ``tuned'' non-X-ray background (NXB) for our HXD/PIN data provided by the HXD team and extracted source and background spectra using the same good time intervals. The PIN spectrum was then corrected for dead time and the exposure time of the background spectrum was increased by a factor 10, as required. Finally, the contribution from the cosmic X-ray background (CXB) was subtracted from the source spectrum, simulating it as suggested by the HXD team. For the sake of simplicity, we will always refer to observation S1, S2, S3 and S4 in this paper, as listed in Table \ref{obslog}, where the final net exposure times for the three XIS spectra and the HXD/PIN are reported. 

\begin{table*}
\caption{\label{obslog}Log for the 2007 XMM-\textit{Newton} and the four \textit{Suzaku} observations.}
\begin{center}
\begin{tabular}{c|cccccc}
\textbf{Obs}& \textbf{Obsid} & \textbf{Date} & $\Delta t$ & \textbf{PN} & \textbf{XIS0}& \textbf{PIN}\\
(1) & (2) & (3) & (4) & (5) & (6) & (7) \\
\hline
XMM07 & 0405380701 & 2007-04-30 & -- & 15 & -- & -- \\
S1 & 702052010 & 2007-05-01 & $<1$ &-- &24 & 20\\
S2 & 702052020 & 2007-05-28 & 27 &-- &29 & 25\\
S3 & 702052030 & 2007-11-09 & 165 &-- &29 & 23\\
S4 & 702052040 & 2007-11-16 & 7 &-- &32 & 24\\
\end{tabular}
\end{center}
(1) The name which identifies the observation in this work; (2) XMM-\textit{Newton} or \textit{Suzaku} observation identifier; (3) Observation date; (4) Time elapsed from previous observation (days); (5) Net exposure time for the EPIC pn (ks); (6) Net exposure time for the XIS0 (ks); (7) Net exposure time for the HXD PIN (ks)

\end{table*}

As a final note, let us discuss the possible contamination of other sources in the field of view (FOV) of the XIS and, most of all, the PIN. The two brightest X-ray sources close to NGC~7582 are the Seyfert 2 galaxy, NGC~7590, and the BL Lac PKS~2316-423. Both are largely outside the extraction regions adopted for the three XIS detectors. As for the PIN, we reconstructed the FOV of each observation, using the tool \textsc{aemkreg} and the actual Euler angle of each pointing. In all the cases, both sources are inside the PIN FOV, but, as already noted by \citet{turn00} for the \textit{BeppoSAX} PDS (whose FOV is larger than that of the PIN), they are significantly softer than NGC~7582 and their hard X-ray fluxes are much dimmer. Moreover, the BL Lac lies very close to the border of the 34 x 34 arcmin square which represents the Full Width Half Maximum of the PIN FOV, thus contaminating only for $\simeq50$ percent of its flux\footnote{See Sect. 8.3 of ``The Suzaku Technical Description''.}.

\subsection{\label{xmm}XMM-\textit{Newton}}

XMM-\textit{Newton} observed NGC~7582 in two targeted exposures on 2001 May 25th and 2005 April 29th. Both observations were discussed in \citet{pico07}. Moreover, the source is within the EPIC field of view of another target, observed on 2007 April 30th, accidentally just a day before the first \textit{Suzaku} one. In this paper, we present for the first time the 2007 observation.
The observation was performed with the EPIC CCD cameras, the pn and the two MOS, operated in Full Window and Medium Filter. Data were reduced with \textsc{SAS} 8.0.0 and screening for intervals of flaring particle background was done consistently with the choice of extraction radii, in an iterative process based on the procedure to maximize the signal-to-noise ratio described by \citet{pico04}.
After this process, the net exposure time was about 15 ks for the pn spectrum, adopting an extraction radius of 19 arcsec and patterns 0 to 4. In the following, we conservatively decided not to use MOS data, which may be affected by possible cross-calibration issues related to the obscuration by the Reflection Grating Array (Mateos et al., in preparation). 
The background spectra were extracted from source-free circular regions with a radius of 50 arcsec. Finally, spectra were binned in order to oversample the instrumental resolution by at least a factor of 3 and to have no less than 25 counts in each background-subtracted spectral channel. The latter requirement allows us to use the $\chi^2$ statistics.

\section{\label{analysis}Spectral analysis}

In the following, errors correspond to the 90\% confidence level for one interesting parameter ($\Delta \chi^2 =2.71$), where not otherwise stated. The adopted cosmological parameters are $H_0=70$ km s$^{-1}$ Mpc$^{-1}$, $\Omega_\Lambda=0.73$ and $\Omega_m=0.27$ \citep[i.e. the default ones in \textsc{xspec 12.4.0}:][]{xspec}. In all the fits, the Galactic column density along the line of sight to NGC~7582 is included \citep[$1.9\times10^{20}$ cm$^{-2}$: ][]{dl90}.

All \textit{Suzaku} XIS instruments were used, but, in order to avoid inter-calibration issues at low energies, the full band (0.5-10 keV) was employed only for the back-illuminated XIS1, which has the largest effective area at low energies, while a restricted band (2-10 keV) was preferred for the two front-illuminated XIS0 and XIS3. We added a normalization constant for each instrument, fixing to 1 the value for XIS0, and to 1.18 that for the PIN (18-50 keV), as appropriate for data taken at the HXD nominal position\footnote{Although the normalization factor we adopted refers to the PIN band of 12-40 keV (see http://www.astro.isas.ac.jp/suzaku/doc/suzakumemo/\\suzakumemo-2008-06.pdf and http://heasarc.gsfc.nasa.gov/docs/suzaku/analysis/\\watchout.html), it was estimated that the variation of this factor when different bands are used for the PIN is roughly less than 2 per cent (see ftp://legacy.gsfc.nasa.gov/suzaku/doc/xrt/suzakumemo-2007-11.pdf), which is lower than the uncertainty on the normalization factor itself (see again http://heasarc.gsfc.nasa.gov/docs/suzaku/analysis/\\watchout.html)}. The values for the XIS1 and XIS3 instruments were left free to vary.

\begin{figure}
\begin{center}
\epsfig{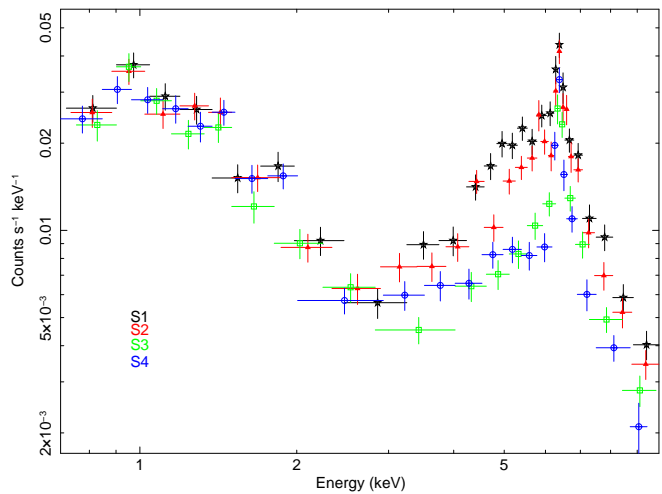}
\end{center}
\caption{\label{allobs}NGC~7582: \textit{Suzaku} XIS0 spectra for the four observations of NGC~7582: S1 (\textit{black}), S2 (\textit{red}), S3 (\textit{green}) and S4 (\textit{blue}).}
\end{figure}

A first look at the \textit{Suzaku} XIS spectra of the four observations immediately reveals that the source varied dramatically above 3 keV (see Fig. \ref{allobs}). On the other hand, the HXD pin spectra, although in a less conclusive way because of the much lower statistics and much larger uncertainty, appear fairly in agreement with each other. Therefore, in the next Sections we will investigate the origin of the variability by fitting separately the available datasets.

\subsection{The `low-flux' state}

We started our analysis from observation S4, which closely resembles the `low-flux' state already observed in the 2005 XMM-\textit{Newton} spectrum. We therefore decided to adopt the model used by \citet{pico07} for the above-mentioned dataset, which, given the longer exposure, has a better signal-to-noise. The model is constituted by an highly obscured powerlaw\footnote{The column densities reported in this paper take into account both photoelectric absorption and Compton scattering (\textsc{cabs*zwabs} in \textsc{Xspec}).} and a pure Compton reflection component (model \textsc{pexrav} in \textsc{Xspec}), together with a Gaussian line to model iron K$\alpha$ emission. The latter two components, though unaffected by the large column density which obscures the primary component, are absorbed by a smaller column density. The soft excess is modeled with a powerlaw and as many Gaussian lines as required, in order to mimic the emission from extended photoionized gas commonly found in Seyfert 2s \cite[see e.g.][]{gb07}. As found for the 2005 XMM-\textit{Newton} observation, this model is highly satisfactory also for the \textit{Suzaku} data of the fourth observation ($\chi^2=91/94$ dof): see last column of Table \ref{bestfits}, where all the best fit parameters are listed.

\begin{table*}
\caption{\label{bestfits}Best fit parameters for the 2007 XMM-\textit{Newton} and the four \textit{Suzaku} observations.}
\begin{center}
\begin{tabular}{cc|ccccc}
&& XMM07 & S1 & S2 & S3 & S4\\
\hline
$\Gamma$ &(1)&  $1.92^*$ & $1.92^*$ & $1.92^*$ & $1.92^*$ & $1.92^{+0.24}_{-0.16}$\\
$\Gamma_\mathrm{r}$ &(2)&  $1.92^*$ & $1.92^*$ & $1.92^*$ & $1.92^*$ & $1.92^{+0.24}_{-0.16}$\\
$\mathrm{N_H^e}$ &(3)&  $6.1^{+3.7}_{-3.3}$ & $1.9^{+1.7}_{-0.8}$ & $<3.0$ & $5.0^{+2.5}_{-2.2}$ & $3.9^{+3.8}_{-3.0}$\\
$\mathrm{N_H^i}$ &(4)&  $33^{+4}_{-5}$ & $44^{+3}_{-2}$ & $68^{+6}_{-7}$ & $110^{+14}_{-11}$ &$120\pm20$\\
$\mathrm{A_{pex}}$ &(5)&  $9.3^*$ & $9.3^*$ & $9.3^*$ & $9.3^*$ & $9.3\pm2.1$\\
$\mathrm{A_{po}}$ &(6)& $10.8^{+1.7}_{-3.3}$ & $8.7^{+0.4}_{-1.0}$ & $11.7^{+2.0}_{-2.2}$ & $17^{+7}_{-5}$ & $14^{+9}_{-6}$\\
$\mathrm{E_{Fe}}$ &(7)&  $6.39^{+0.05}_{-0.03}$ & $6.419^{+0.018}_{-0.021}$ & $6.408^{+0.016}_{-0.018}$ & $6.421\pm0.018$ & $6.411^{+0.013}_{-0.014}$\\
$\mathrm{\sigma_{Fe}}$ &(8)&  $<170$ & $50^{+30}_{-40}$ & $<70$ & $<80$ & $<60$\\
$\mathrm{F_{Fe}}$ &(9)&  $2.4\pm0.9$ & $2.5\pm0.5$ & $2.2\pm0.4$ & $2.2\pm0.4$ & $2.4\pm0.3$\\
$\mathrm{F_{0.5-2}}$ &(10)&  $0.32\pm0.04$ & $0.34\pm0.08$ & $0.40\pm0.13$ & $0.42\pm0.09$ & $0.43\pm0.08$\\
$\mathrm{F_{2-10}}$ &(11)&  $7.6\pm0.4$ & $5.3\pm0.3$ & $4.1\pm0.3$ & $3.2\pm0.2$ & $2.6\pm0.5$\\
$\mathrm{L_{2-10}}$ &(12)&  $2.0\pm0.1$ & $1.5\pm0.1$ & $2.1\pm0.4$ & $3.2\pm1.1$ & $2.9\pm1.4$\\
$\chi^2$/dof &(13)&  87/83 & 114/110 & 112/111 & 85/81 & 91/94\\
\end{tabular}
\end{center}
$^*$ Fixed. 
(1) Primary continuum powerlaw index - (2) Reprocessed emission powerlaw index (Compton reflection component and soft X-ray emission) - (3) External column density ($10^{22}$ cm$^{-2}$) - (4) Internal column density ($10^{22}$ cm$^{-2}$) - (5) Compton reflection component normalization ($10^{-3}$ ph keV$^{-1}$ cm$^{-2}$ s$^{-1}$) - (6) Primary powerlaw normalization ($10^{-3}$ ph keV$^{-1}$ cm$^{-2}$ s$^{-1}$) - (7) Iron K$\alpha$ emission line centroid energy (keV) - (8) Iron K$\alpha$ emission line physical width (eV) - (9) Iron K$\alpha$ emission line flux ($10^{-5}$ ph cm$^{-2}$ s$^{-1}$) - (10) Observed 0.5-2 keV flux ($10^{-12}$ erg cm$^{-2}$ s$^{-1}$) - (11) Observed 2-10 keV flux ($10^{-12}$ erg cm$^{-2}$ s$^{-1}$) - (12) Unabsorbed 2-10 keV luminosity ($10^{42}$ erg s$^{-1}$) - (13) Best fit $\chi^2$/dof.
\end{table*}

A comparison with the results of the 2005 XMM-\textit{Newton} observation is very interesting. We recover the same best fit parameters reported in \citet{pico07}, such as the photon index and no significant variation is detected for the two absorbing column densities. Moreover, the normalizations of the Compton reflection component are perfectly consistent between the 2005 XMM-\textit{Newton} observation ($9.7^{+0.8}_{-1.8}\times10^{-3}$ ph keV$^{-1}$ cm$^{-2}$ s$^{-1}$)\footnote{This value was not explicitly reported in \citet{pico07} and was found re-analyzing the 2005 XMM-\textit{Newton} observation. All the other parameters are perfectly consistent with the one published by \citet{pico07}.} and the \textit{Suzaku} data. The neutral iron K$\alpha$ line flux does not show any significant variability between the two datasets, either. Its equivalent width (EW) with respect to the Compton reflection component is around 1 keV, as expected if arising as reprocessing from the same Compton-thick material \citep[e.g.][]{mpp91}. Finally, while we refer the reader to \citet{pico07} for a detailed analysis of the soft X-ray emission of NGC~7582, we confirm here the detection of strong K$\alpha$ lines from {Ne\,\textsc{IX}}, {Ne\,\textsc{X}}, {Mg\,\textsc{XI}} and {Si\,\textsc{XIII}}, as well as from {Fe\,\textsc{XVII}} L. The observed 0.5-2 keV flux is around 1.5 per cent of the nuclear one, after correction for absorption, in the range usually found for Seyfert 2 galaxies \citep{bg07}. 

\subsection{The origin of the variability}

The results in the previous section strongly suggest that some important X-ray parameters in NGC~7582 remain constant over long periods of time. Considering also the 2001 XMM-\textit{Newton} observation, the intrinsic powerlaw index, the normalization of the Compton reflection component and the flux of the iron K$\alpha$ line do not show significant variability in 6 years. On the other hand, it is clear from the two XMM-\textit{Newton} and the 4 \textit{Suzaku} observations that the spectrum has dramatically varied. What is the main driver of this variability?

The complexity of the best fit model implies some degeneracy between the spectral parameters, most of all in the datasets where the presence of a stronger primary continuum does not allow us to disentangle unambiguously the Compton reflection component. However, we can now reasonably assume that the powerlaw index and the Compton reflection components are not affected by variability. Therefore, in all the following fits we decided to use the photon index and the \textsc{pexrav} normalization fixed to the best fit values found in S4. We leave the neutral iron K$\alpha$ line flux and centroid energy free to vary, since it is much easier to detect variability in this sharp feature with respect to a continuum component. This is a consistency test: given the common origin of the iron K$\alpha$ line and the reflection component in our model, any variation of the former would invalidate our assumption.

\begin{figure*}
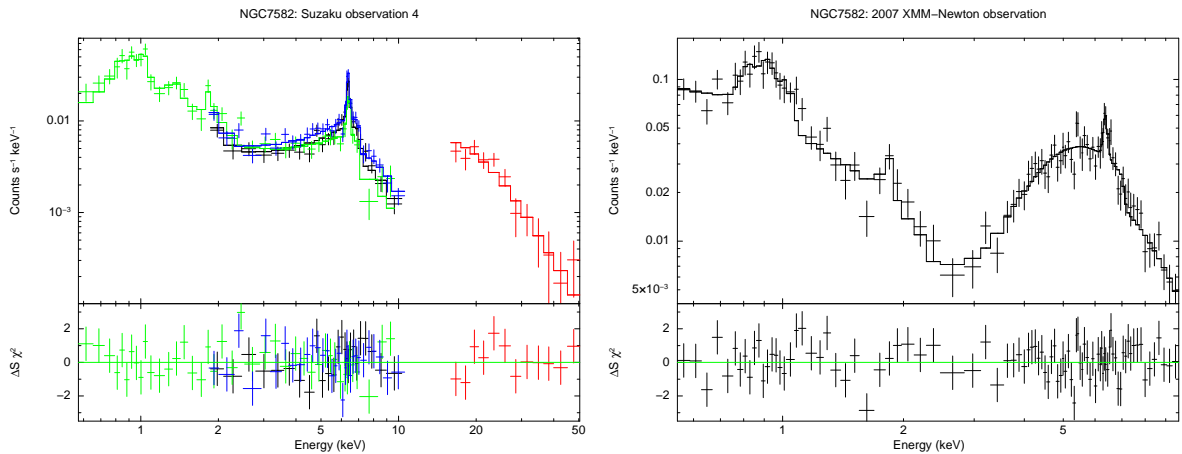

\begin{center}
\epsfig{file=suzaku4.ps, width=6.cm,angle=-90}
\epsfig{file=xmmnew.ps, width=6.cm,angle=-90}
\end{center}
\caption{\label{bestplots}NGC~7582: \textit{Suzaku} observation 4 (\textit{Left}) and 2007 XMM-\textit{Newton} observation (\textit{Right}) data and best fit model.}
\end{figure*}

Adopting the model described above, we then fitted the other three \textit{Suzaku} spectra and the new 2007 XMM-\textit{Newton} data. In all cases, we found very good fits (see Table \ref{bestfits} and Fig. \ref{bestplots}). Allowing the frozen parameters to vary does not significantly improve the fits (whose reduced $\chi^2$ are already very close to unity), but only enlarges the statistical uncertainties on the parameters. All the main parameters of the model appears to be constant among the observations. In particular, the neutral iron K$\alpha$ line does not show any significant variation, thus being consistent with our assumption that the reprocessed components from Compton-thick material are indeed constant. As expected, no significant variability is found in the flux and modelization of the soft X-ray emission, either.

\begin{figure}
\begin{center}
\epsfig{file=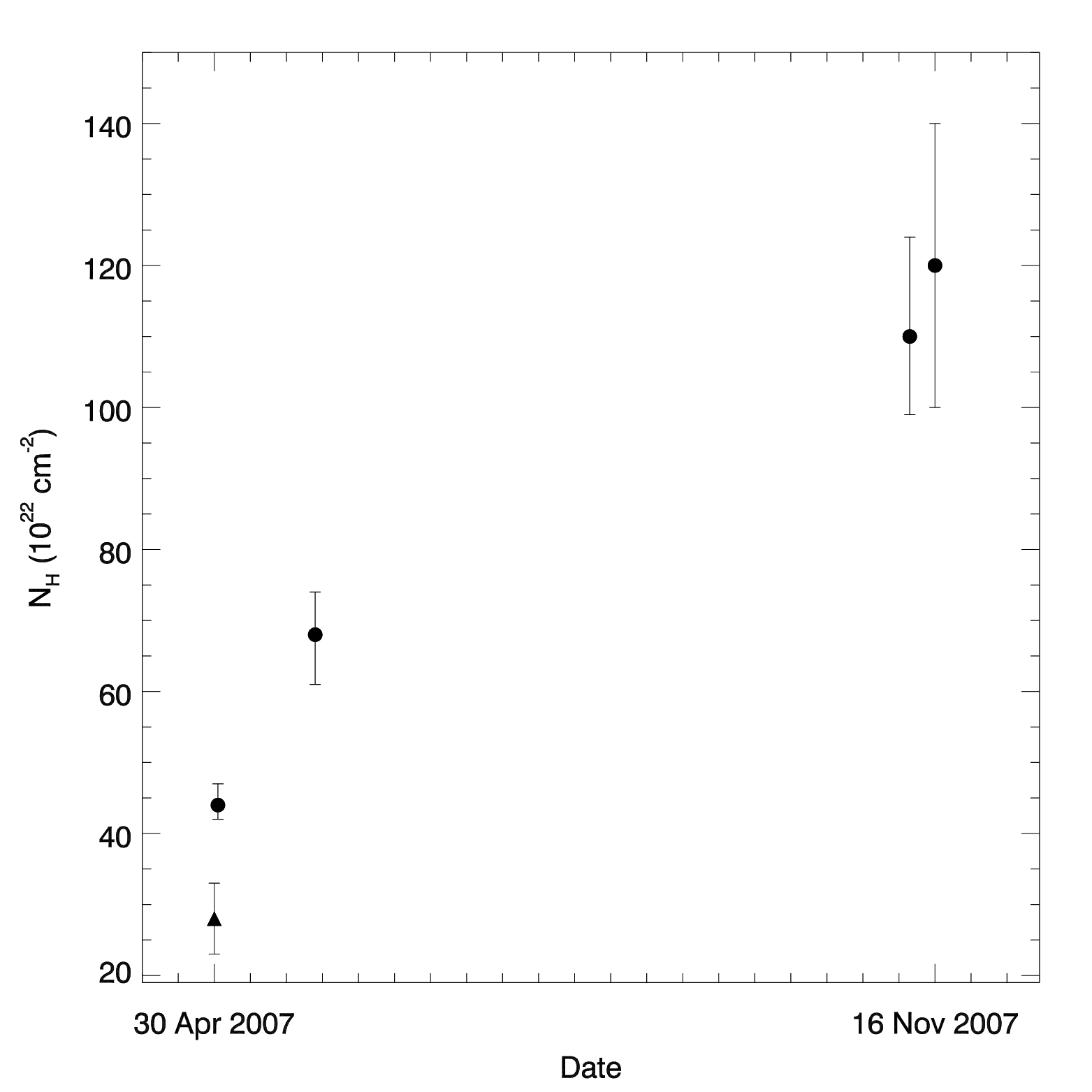, width=8cm}
\end{center}
\caption{\label{nhplot}NGC~7582: Column density of the inner absorber in the four \textit{Suzaku} observations (circles) and the latest XMM-\textit{Newton} one (triangle). Time bins on the abscissa are 10 days long.}
\end{figure}

Therefore, the clear variability observed among the \textit{Suzaku} observations has to be ascribed to the behavior of the inner column density, the only parameter which significantly changes between the observations (see Table \ref{bestfits} and Fig. \ref{nhplot}). The variation (from $7\times10^{23}$ to $1.1\times10^{24}$ cm$^{-2}$) between S2 and S3 occurs in roughly 5 months. A much shorter timescale, the 23 days separating the first from the second observation, witnesses another significant variation, from around $4.5$ to $7\times10^{23}$ cm$^{-2}$. A still shorter timescale (less than a day) characterizes the variation from $3.3^{+0.4}_{-0.5}$ to $4.4^{+0.3}_{-0.2}\times10^{23}$ cm$^{-2}$ measured between XMM07 and S1. In conclusion, our best fit model allows us to ascribe most of the observed spectral variability in NGC~7582 to the rapid changes of the column density of this internal absorber. On the other hand, there are some hints of variability of the primary continuum intensity, but they are not conclusive, given the large errors.

\section{Discussion}

\subsection{NGC~7582: the big picture}

The fourth and latest \textit{Suzaku} observation caught the source at the lowest state, but this indeed allowed us to have a clearer view of the reprocessing components of its spectrum, which apparently are those of a typical Seyfert 2. The resulting scenario applies well to all other X-ray observations of NGC~7582, the different states being due only to the variability of the column density of the inner absorber. In this section, we will discuss in detail the implications on the complex geometry of the absorbers required in this source.

The intrinsic nuclear emission appears obscured by a very large column density (just below the `canonical' Compton-thick limit). The spectrum below 10 keV is therefore dominated by a Compton reflection component and the relative iron line. Both the flux of the reflection component and of the iron line are consistent with being constant during the \textit{Suzaku} monitoring campaign and with the values found in XMM-\textit{Newton} observations, the older dating back to 2001. The material that produces both components is likely to be quite far away from the nuclear X-ray emitting source, possibly in the classic pc-scale `torus' invoked in Unification Models \citep{antonucci93}.

These reprocessing components appears to be obscured by a second absorber, which must be located farther away. Its column density is not well constrained in the \textit{Suzaku} spectra, but is consistent with the one measured by XMM-\textit{Newton}, around $4-5\times10^{22}$ cm$^{-2}$. It can be identified with a large scale obscuration, as the dust lanes commonly observed in galaxies. Indeed, the combined analysis of \textit{HST} and \textit{Chandra} images clearly detected such a dust lane also in the X-rays, with a column density consistent with the one required by the spectral fits \citep{bianchi07b}. In this case, the presence of a second, Compton-thin absorber, as invoked in simple modifications of the Unified Models \citep[as in][and references therein]{matt00b}, is directly observed.

The soft X-ray emission, as reported by \citet{pico07} thanks to the well-exposed XMM-\textit{Newton} Reflection Grating Spectrometer (RGS) high resolution spectra, appears dominated by emission lines of highly ionized species. This is a general characteristic of Seyfert 2 galaxies, as found by \citet{gb07}. The lack of any variability of the soft X-ray emission is in agreement with the scenario, where these emission lines are produced in a large scale material, spatially coincident with the Narrow Line Emission (NLR) and likely dominated by photoionization from the AGN \citep[e.g.][]{bianchi06}.

However, our monitoring campaign discovered a striking feature that characterizes NGC~7582: the presence of an absorber, whose rapid variability imposes a location far closer to the BH than the torus. While the most rapid variation occurs between XMM07 and S1 (less than a day), significant variability is also observed on larger timescales between the \textit{Suzaku} observations. The distance of the absorber can be roughly estimated with the following reasoning, based on \citet{risa02b}. We can assume, for simplicity, that the absorbing material is constituted of individual spherical clouds with column density $N_\mathrm{H}$ and density $n$. In this scenario, the variability in the column density that we are observing is due to the complete passage of a given cloud of dimension $D\simeq N_\mathrm{H}/n$. The time required for this cloud to pass completely off the line of sight is not larger than $t=D/v$. Assuming that the clouds are rotating at a distance $R$ from the BH with Keplerian velocities, this relation translates into $t=D\left(GM_\mathrm{BH}/R\right)^{-\frac{1}{2}}$. Therefore:

\begin{align}
\label{distance}
R&=\frac{GM_\mathrm{BH}t^2n^2}{N_\mathrm{H}^2}\nonumber\\
&\simeq4\times10^{15} \left(\frac{M_\mathrm{BH}}{5.5\times10^7\,M_\odot}\right) \left(\frac{t}{20\,\mathrm{h}}\right)^2\nonumber\\ &\left(\frac{n}{10^{9}\,\mathrm{cm}^{-3}}\right)^2 \left(\frac{N_\mathrm{H}}{10^{23}\,\mathrm{cm}^{-2}}\right)^{-2} \, \mathrm{cm}
\end{align}

\noindent where we have normalized all the parameters to our fiducial values for the case we are discussing: the BH mass was estimated by \citet{wold06} as $5.5\times10^7\,M_\odot$, $N_\mathrm{H}\simeq10^{23}$ cm$^{-2}$ is the difference observed between XMM07 and S1, $t\simeq20$ hours is the time elapsed between these two observations. A crucial value is the electron density: the value $n=10^{9}$ cm$^{-3}$ corresponds to a cloud dimension $D=1\times10^{14}$ cm, i.e. roughly 10 $r_g$ for the BH mass estimate reported above. A larger density would shift the material to larger distances, but implying dimensions of the clouds smaller than 10 $r_g$ and, therefore, likely smaller than the X-ray source they should obscure. On the other hand, lower densities would result in a still smaller distance, which is already an upper limit, given that the actual crossing time must be lower than the separation between the two observations. There is a physical limit at this distance, which cannot be smaller than the dimension of the clouds $D$. This gives a lower limit for the density, which is $n>3\times10^8$ cm$^{-3}$.

On the basis of the average unabsorbed X-ray luminosity of NGC~7582, as derived in this work ($<L_{2-10\,keV}>\simeq2.3\times10^{42}$ erg cm$^{-2}$ s$^{-1}$), we can estimate the radius of the BLR in this source to be around $R_{BLR}=0.5-1\times10^{15}$ cm \citep[see][for the relation between the two parameters and the relative uncertainties]{kaspi05}. Moreover, a typical density for the BLR is believed to be $10^{9.5}$ cm$^{-3}$, or larger for the inner regions \citep[e.g.][]{peterson97}. Therefore, the location and physical properties of the absorbing clouds in NGC~7582 are consistent with being within or immediately outside the BLR. Note that the sublimation radius in this source is beyond $10^{17}$ cm \citep[see e.g.][]{barv87}, thus the BLR and the X-ray absorbing clouds must be dust-free. This means that these clouds are not responsible for the absorption of the BLR and the consequent classification of NGC~7582 as a Seyfert 2. This is due to the large-scale absorber with column density of a few $10^{22}$ cm$^{-2}$. On the other hand, we stress again that the torus is not along the line of sight in this source. This scenario will be generalized in the next Section.

\subsection{Do we need a new Unification Model?}

X-ray observations have collected much evidence in favor of the presence of the pc-scale \textit{torus} envisaged in Unification Models. In particular, the ubiquitous presence of a Compton reflection component, invariably accompanied by a neutral iron narrow K$\alpha$ emission line, is a clear signature of the presence of Compton-thick material also in Type 1 objects, even if it does not intercept the line of sight \citep[see e.g.][]{per02,bianchi04,bianchi07}. The distance of the \textit{torus} is generally inferred from the lack of variability of these components between observations separated years one from the other. Therefore, a pc-scale \textit{torus} must be an fundamental ingredient of any Unification Model. 

However, there is now a growing number of sources which show dramatic absorption variability in timescales as short as hours \citep[e.g.][]{elvis04,ris05,puc07}. These objects cannot be described in the framework of classic Unification Models and they are generally considered exceptions of an otherwise successful scenario. However, it is likely that much more sources would present the same characteristics, if only they were observed with aimed monitoring campaigns, as we did for NGC~7582.

We propose that the simplest scenario that fits the X-ray observations should consider the presence of three neutral absorbers/emitters, even if not necessarily coexisting or observable in all the sources. A Compton-thick torus is likely present in the vast majority of the sources at a distance from the BH roughly around a pc. It is responsible for the production of the Compton reflection component and the neutral iron narrow K$\alpha$ line, both invariably present in all X-ray spectra of AGN and generally found not to vary on timescales shorter than years. If the torus intercepts the line of sight, the observer classifies the object as a Compton-thick Seyfert 2.

On a much larger scale, a Compton-thin absorber with column density around $10^{22}$ cm$^{-2}$ may intercept the line of sight, completely or partially obscuring the BLR in the optical and absorbing the X-ray spectrum. The effect of this material, likely associated to dust lanes, is to force the observer to classify the object as an intermediate Seyfert 1 or a Compton-thin Seyfert 2.

To this dual-absorber scenario, basically the same proposed by \citet{matt00b}, a third material should be added, on a scale much shorter than the torus, roughly where the BLR is located. This material cannot be seen in Compton-thick Seyfert 2s, i.e. those sources absorbed by the torus, because it is obscured by the torus itself. It is responsible for fast variability of the absorbing column density. From an observational point of view, this material can be effectively discriminated from the torus if it is patchy. In this case, given the close distance to the BH, the chance to see a cloud appearing and disappearing along the line of sight is not low, on short timescales. In the exceptional case of NGC~1365, a clear case of eclipse from a cloud is actually observed \citep{ris07}. If the cloud is not Compton-thick, you may still observe Compton-thin Seyfert 2 with large column densities (as large as several $10^{23}$ cm$^{-2}$, for example), likely varying on short timescales.

It is important to stress that such a material does not relax our need for a torus. The latter is needed because nearly all the observed AGN have a Compton reflection component and neutral iron K$\alpha$ line, which do not show significant variability up to quite long timescales. In order to reproduce this observational evidence, the material must be Compton-thick, with large covering factor and, most of all, quite far from the BH, unless the nuclear emission remains constant (but this is not the case for many AGN). Only a pc-scale torus has all these characteristics.

This scenario (summarized in Table \ref{unifmodel}) makes a number of predictions. Most of the Compton-thick Seyfert 2s are likely still absorbed by the torus and are not expected to show any flux or spectral change on timescales lower than years. However, a fraction of Compton-thick objects does not intercept the torus along the line of sight, but are caught when a Compton-thick cloud located in the BLR is passing in front of the source. Such sources may change their status in a following observation, once the cloud has passed, explaining some of the so-called `changing-look' objects \citep[e.g.][the alternative being a `switching-off' of the nucleus]{mgm03,gua05,bianchi05c,teng08} and, definitely, NGC~1365 \citep{ris07}. The fraction of Compton-thick sources belonging to the two classes basically depends on the covering factors of the torus and the inner absorber.

\begin{table*}
\caption{\label{unifmodel}A new Unification Model, based on three absorbers, located at different distances from the BH. Their presence along the line of sight (highlighted by a $\surd$) determines the classification of the object. As for the torus, it is required in all cases, because of the ubiquity of not-variable reprocessed components from Compton-thick material, even if only Compton-thick sources intercepts it along the line of sight. See text for details.}
\begin{center}
\begin{tabular}{c|ccc}
\textbf{Classification}& \textbf{Dust lane} & \textbf{Torus} & \textbf{Clouds}\\
& ($>>$ pc) & (pc) & ($<$ pc)\\
\hline
Seyfert 1 & & &\\
`Changing-look' Seyfert 1 & & &$\surd$\\
Compton-thin Seyfert 2 & $\surd$ & &\\
`Changing-look' Seyfert 2 & $\surd$ & &$\surd$\\
Compton-thick Seyfert 2 & ? & $\surd$ & ?\\
\end{tabular}
\end{center}

\end{table*}

Most of the Compton-thin Seyfert 2s with column densities of the order of $10^{22}$ cm$^{-2}$ do not intercept at all the torus along the line of sight, but are absorbed by large scale dust lanes, which are also responsible for the obscuration of the optical broad emission lines. On the other hand, Seyfert 2s with larger column densities, of the order of $10^{23}$ cm$^{-2}$, are likely seen through the absorbing clouds located at the BLR, in analogy to the `changing-look' objects cited above, the only difference being that the intervening clouds are not Compton-thick. These sources are probably the best candidates for monitoring campaigns, since they are those with higher probability of rapid column density variations. NGC~7582 is a clear example of this class. Again, the fraction of these sources among Compton-thin Seyfert 2s depends on the covering factor and the geometry of the inner absorber.

Finally, Seyfert 1s are those objects where none of the three materials intercepts the line of sight. However, depending on the geometry and covering factor of the inner absorber, they may occasionally show even rapid partial or total occultation events. Indeed, this is the case of NGC~4051 \citep{gua98}, NGC~3227 \citep{lamer03} Mrk~335 \citep{gkg07} and H0557-385 \citep{long08}. While in some cases an alternative solution in terms of `switching-off' of the source is equally viable, in the last two cases the partial covering from an intervening absorber is preferred. Even if the timescales are not particularly constraining, the presence of warm absorption signatures unchanged between the two states of H0557-385 strongly suggests that the neutral absorbing clouds are located close to the BH, in agreement with the predictions of the model proposed in this paper. We note here that simultaneous optical and X-ray campaigns of these sources would be revealing for the exact location of the absorber with respect to the BLR, looking for the presence of broad optical lines during the X-ray absorption states.

While the study of the nature of this inner absorber is well beyond the scope of this paper, we note here that this scenario is well in agreement with theoretical models which suggest a strong link, both geometrical and physical, between the accretion disk and the BLR, and possibly the torus itself \citep[see e.g.][]{elvis00,nic00,es06,elvis06}. In any case, we would like to point out that our simplification in terms of two separate materials, one compact, roughly at a pc from the BH and Compton-thick (the torus) and the other close to the BH and constituted by Compton-thick and/or thin clouds, can reproduce the observed phenomenology.

\section{Conclusions}

We have presented a \textit{Suzaku} monitoring campaign of the Seyfert 2 galaxy, NGC~7582. The dramatic spectral variability observed during the 4 observations is best explained by changes of the absorbing column density of a material close to the X-ray primary source. Given the significant variation between a new XMM-\textit{Newton} observation and the first \textit{Suzaku} one, separated by only 20 hours, its distance can be estimated to be a few $\times10^{15}$ cm, i.e. roughly consistent with the BLR.

Together with this material, the presence of a Compton-thick material on larger scale, likely the `torus' envisaged in the Unification models, is required in order to account for the Compton reflection component and the neutral iron K$\alpha$ emission line, whose fluxes appear constant in years. On the top of that, a third, Compton-thin material intercepts the line of sight and can be associated to the dust lane observed in optical and \textit{Chandra} X-ray images.

In the last years, a number of sources have shown a geometry for the absorbers which is necessarily more complex than what generally assumed in simple Unification Models. NGC~7582 is a clear-cut example, where three neutral absorber/emitter regions must be present on very different scales. We propose that the scenario adopted for this source may be the rule rather than the exception. The Unification Model should therefore be modified in order to account for all the observational evidence collected so far.

The new scenario is based on the model presented by \citet{matt00b}: to the ubiquitous pc-scale torus, an extended, Compton-thin material, likely associated to galactic dust lanes, has to be added. Moreover, we suggest the presence of another material, made of Compton-thick and/or Compton-thin clouds, located roughly at the BLR, whose presence can be unveiled only when the torus does not intercept the line of sight and the material is patchy, leading to observable absorption variability on short timescales.

\section*{Acknowledgements}

SB and GM acknowledge financial support from ASI (grant 1/023/05/0). We would like to thank the anonymous referee for helpful suggestions.

\bibliographystyle{apj}
\bibliography{sbs}

\end{document}